\documentstyle[12pt]{article}
\begin{document}
\newcommand{\be}{\begin{equation}}
\newcommand{\ee}{\end{equation}}
\newcommand{\bq}{\begin{eqnarray}}
\newcommand{\eq}{\end{eqnarray}}

\pagestyle{empty}
\begin{flushright}
{CEBAF-TH-93-02\\
January 1993}
\end{flushright}
\vspace{.4cm}
\begin{center}

{\Large \bf Delta-Isobar Magnetic Form Factor in QCD }
\vspace{1 cm}

{\large  V.M.Belyaev} \\
\vspace {0.3 cm}
{\it Institute of Theoretical and Experimental Physics\\
 B.Cheremushkinskaya 25,\\ 117259, Moscow, USSR \\
and\\
Continuous Electron Beam Accelerator Facility\\
12000 Jefferson Ave, Newport News, Virginia 23606, USA\\}
\vspace{0.6 cm}

 \end{center}
\vspace{0.4cm} \noindent

\begin{abstract}

We consider the  QCD sum rules approach for $\Delta$-isobar
magnetic
 form factor in
the infrared region
$0 \leq Q^{2} < 1 GeV^{2}$.   The QCD sum rules in external
 variable field
 are used.
The obtained form factor
 is in  agreement with quark model predictions for the $\Delta$-isobar
magnetic moment.

\end{abstract}

\newpage
\pagestyle{plain}
\setcounter{page}{1}

\section{Introduction}

The QCD sum rule method suggested  by Shifman, Vainshtein and
 Zakharov (SVZ) in the pioneering paper \cite{SVZ} becomes
 now a universal tool for calculating different properties of low-lying
hadronic
states. Using the original version of this method, the meson \cite{SVZ}
 and baryon \cite{baryon}  masses were found from the sum rules for
two-point correlation functions. Using the three-point correlation functions,
 hadron
form factors at intermediate $Q^{2}$ can be obtained
\cite{3}. Unfortunately, this method does not work if one tries to calculate
 form factors in the infrared region $0<Q^{2}<1 GeV^{2}$ due
to  power corrections $1/Q^{2n}$  at
$Q=0$. The new method - QCD sum rules in
external field was suggested in \cite{ext}, and
using this method nucleon magnetic moments were found \cite{mom} as well as
baryon axial couplings \cite{kobel}. Then this method was formulated
 for a variable external field \cite{fn}
which gives a possibility to calculate form factors
at $Q^{2} \neq 0$.

In \cite{fn} we have formulated a new method  for calculating
hadronic form factors in the infrared region.
 To study a form factor at
nonzero $Q^{2}$, it is necessary to introduce a variable external field.
   The calculation of a polarization operator in
this field encounters a number of difficulties as compared with the case of a
constant external field. The arising problems and methods to avoid them
   were
discussed in detail in the paper \cite{fn}.

 Let us note that in  the papers \cite{radius} and \cite{rn},
 the pion and nucleon
 charge radii were considered, using the methods similar to ours.
However, the results obtained in \cite{radius} are  connected  not
 with the calculation
 of
 the total form factor  $F(Q^{2})$, but only the  first derivative at zero
momentum transfer $<r^{2}> \sim F'(Q^{2})|_{Q=0}$.
In \cite{rn} pion form factor was considered.
The aim of the paper is to use the general method for calculating
$\Delta$-isobar magnetic
 form factor in the infrared region.

\section{Polarization Operator in Variable External Field}

  To compute $\Delta$-isobar magnetic form factor we shall consider
the following correlator in an external variable electromagnetic
field:
\be
\Pi_{\mu\nu}^{V*}(p,k) = i\int 2\pi\delta (kx)
e^{ipx} d^{4}x  < T\{ \eta_\mu (x), \bar{\eta}_\nu (0)\} >_{V}
\label{1}
\ee
$$
=\int_{-\infty}^{+\infty}dzi\int e^{i(p+kz)x}d^4x
< T\{ \eta_\mu (x), \bar{\eta}_\nu (0)\} >_{V}
$$
$$
=\int_{-\infty}^{+\infty}dz\Pi_{\mu\nu}^V(p+kz,k)
$$
where
\be
\Pi_{\mu\nu}^{V}(p,k) = i\int
e^{ipx} d^{4}x  < T\{ \eta_\mu (x), \bar{\eta}_\nu (0)\} >_{V}
\label{pmn}
\ee
\be
\eta_{\mu} (x) =
\epsilon^{abc} (u^{a}C\gamma_\mu u^{b})u^{c}
\label{eta}
\ee
is the quark current with the $\Delta$-isobar
 quantum numbers suggested in the first
paper in Ref.\cite{baryon}, $u$ is the $u$-quark  operator;
$a$ , $b$ and $c$ are the color indeces, $\epsilon^{abc}$ is the antisymmetric
tensor and $C = -C^{T}$ is the charge conjugation matrix; index $V$ means
the vacuum average in the presence of weak external electromagnetic
field that is responsible for adding to Lagrangian of the following
term
\be
\Delta {\cal L} = - V_{\mu}e^{ikx}\sum_{f} e_{f}\bar{q}_{f}(x)\gamma_{\mu}
q_{f}(x)=-V_\mu j_\mu (x)e^{ikx}
\label{3}
\ee
where $e_{f}$ is the charge of the quark with flavor $f$,  $V_{\mu}$ and
$k_{\mu}$ are the amplitude and the momentum of the classical external field.
This correlator was suggested in \cite{fn}.

Now let us discuss the reason why we need to introduce the $\delta$-function
in the correlator (\ref{1}). To calculate the correlator (\ref{pmn})
at $p^2\sim -1GeV^2$,
$(p+q)^2\sim -1GeV^2$, $k^2=-Q^2<0)$
we use operator product expansion in the presence of external variable field
(\ref{3}).
So  we need to know nonperturbative quark propagator in the field
\be
<:T\{ q_\alpha^a(x),q_\beta^b(0)\} :>_V
\label{4}
\ee
$$
=<T\{ q_\alpha^a(x),q_\beta^b(0)\} >_V
-<T\{ q_\alpha^a(x),q_\beta^b(0)\} >_V^{(pert.)}
$$

Where $::$ denotes a subtraction of perturbative contribution.
It is possible to find perturbative part of this propagator in
the form of expansion over the coupling constant.  To take
into account nonperturbative
interaction of the quark
with the external field, we expand eq.(\ref{4}) over $x_\mu$
\be
<:T\{ q_\alpha^a(x),q_\beta^b(0)\} :>_V=
<: q_\alpha^a(0),q_\beta^b(0) :>_V
\label{5}
\ee
$$
+x_\mu <: D_\mu q_\alpha^a(0),q_\beta^b(0) :>_V
+\frac{1}{2}x_\mu x_\nu
<:D_\mu D_\nu q_\alpha^a(0),q_\beta^b(0) :>_V
+...
$$

It is clear that the $n$'th term of expansion (\ref{5}) can
give dimensionless
factor $(kx)^n$. Effectively it means that highest terms contribution
of
expansion (\ref{5})
into a polarization operator (\ref{pmn})
is not suppressed  because this factor
$(kx)^n$ corresponds to the factor $(kp)/p^2\sim 1$.
 To kill the dangerous contributions of the terms $\sim (kx)^n$
we insert $\delta (kx)$ into the correlator (\ref{1}).
This correlator (\ref{1}) can be calculated in a form of series over
$1/p^2$.
Therefore at respectively large $-p^2$ this correlator can be
calculated with a good accuracy using only the first few terms in
expansion (\ref{5}).

The nonperturbative quark propagator in the external field has
the following form:
$$
<:u_{\alpha}^a(x), \bar{u}_{\beta}^b(0):> =
e_u\frac{\delta^{ab}}{12}\{ \hat{V}_{\alpha \beta}k^2\Pi_1(k^2)
$$
\be
+(\sigma_{\rho \lambda})_{\alpha \beta}k_\rho V_\lambda \Pi_{2}(k^2) +
(Vx)[i<:\bar{\psi}\psi:>_0 + \frac{1}{2}k^2\Pi_2(k^2)]\delta_{\alpha \beta}
\label{6}
\ee
$$
+\frac{i}{2}\hat{V}_{\alpha \beta}(kx)k^2\Pi_1(k^2) +
\frac{i}{2}(\sigma_{\rho \lambda})_{\alpha \beta}k_\rho V_{\lambda}(kx)
\Pi_2(k^2)
$$
$$
-\frac{1}{4}x_\mu \varepsilon_{\mu \nu \rho
\lambda}(\gamma_\lambda\gamma_5)_{\alpha \beta}
k_\nu V_\rho k^2\Pi_1(k^2)
$$
$$
+\frac{i}{12}x^2(\sigma_{\rho \lambda})_{\alpha \beta}k_\rho V_\lambda
[<:\bar{\psi}\psi :>_0 - i\Pi_1^G(k^2) - 2\Pi_2^G(k^2) + ik^2\Pi_4(k^2)]
$$
$$
+(Vx)(kx)k^2\Pi_3(k^2)\delta_{\alpha \beta} +
\frac{(kx)^2}{2}
(\sigma_{\rho \lambda})_{\alpha \beta}k_\rho V_\lambda \Pi_4(k^2)
$$
$$
+\frac{(kx)}{12}x_\mu (\sigma_{\mu \nu})_{\alpha \beta}V_\nu
[\frac{5}{2}i<:\bar{\psi}\psi :>_0 + \Pi_1^G(k^2) + 3ik^2\Pi_3(k^2) +
i\Pi_2^G(k^2)
$$
$$
-\frac{5}{2}k^2\Pi_4(k^2)] +
\frac{(Vx)}{12}x_\mu (\sigma_{\mu \nu})_{\alpha \beta}k_\nu
[\frac{i}{2}<:\bar{\psi}\psi :>_0 - \Pi_1^G(k^2)
$$
$$
-i\Pi_2^G(k^2) + 3ik^2\Pi_3(k^2) - \frac{k^2}{2}\Pi_4(k^2)]
$$
$$
+\; (terms \; with\; an\; odd\; number\;of\;\gamma -matrices)\} +
O(x^3)
$$
The correlators $\Pi_i(k^2)$ are defined as follows:
\bq
e_{u}\Pi_{1}(k^{2})(k^{2}g_{\mu \nu} - k_{\mu}k_{\nu}) =
\nonumber
\\
i\int e^{ikx}d^{4}x<:T\{ \sum_{f}\bar{q}_{f}\gamma_{\mu}q_{f}(x),
\bar{u}\gamma_{\nu}u(0)\} :>_{0}
\nonumber
\\
e_{u}\Pi_{1}(k^{2})(k_{\mu}g_{\nu \rho} - k_{\nu}g_{\mu \rho}) =
\label{7}
\\
i\int e^{ikx}d^{4}x<:T\{ \sum_{f}\bar{q}_{f}\gamma_{\rho}q_{f}(x),
\bar{u}\sigma_{\mu \nu}u(0)\} :>_{0}
\nonumber
\\
e_u\Pi_3(k^2)(k^2(g_{\mu \rho}k_{\nu} + g_{\nu \rho}k_{\mu}) -
2k_{\mu}k_{\nu}k_{\rho}) =
\nonumber
\\
i\int e^{ikx}d^4x<:T\{ \sum_{f}e_f\bar{q}_f\gamma_{\rho}q_f(x),
\bar{u}\stackrel{\rightarrow}{D_{\{ \mu}D_{\nu \} }}u(0)\} :>_0
\nonumber
\\
e_u\Pi_4(k^2)k_{\mu}k_{\nu}(g_{\rho r}k_p - g_{\rho p}k_r) +... =
\nonumber
\\
i\int e^{ikx}d^4x<:T\{ \sum_{f}e_f\bar{q}_f\gamma_{\rho}q_f(x),
\bar{u}\sigma_{pr}\stackrel{\rightarrow}{D_{\{ \mu}D_{\nu \} }}u(0)\} :>_0
\nonumber
\\
e_u \Pi_1^G (k^2 )(k_{\mu}g_{\rho \nu} - k_{\nu}g_{\rho \mu}) =
\nonumber
\\
i\int e^{ikx}d^{4}x
<:T\{ \sum_{f}e_f \bar{q}_f \gamma_{\rho}q_f (x),
g_s\bar{u}G_{\mu \nu}^{n}t^n u(0)\} :>_0
\nonumber
\\
e_u \Pi_2^G (k^2 )(k_{\mu}g_{\rho \nu} - k_{\nu}g_{\rho \mu}) =
\nonumber
\\
i\int e^{ikx}d^{4}x
<:T\{ \sum_{f}e_f \bar{q}_f \gamma_{\rho}q_f (x),
g_s\bar{u}\tilde{G}_{\mu \nu}^{n}t^n \gamma_5 u(0)\} :>_0
\nonumber
\eq
Perturbative contributions are subtracted in correlators
(\ref{7}). These expressions were obtained in \cite{fn}.

 In this paper we neglect operators $\Pi^G(k^2)$ because  their
 contribution into a sum rule is small (see \cite{fn}).

\section{The Sum Rules}

In this Section, we obtain a sum rule for the $\Delta$-isobar
 magnetic form factor.
First, we should choose a tensor
 structure which has a contribution from the
magnetic transition between baryon states with quantum numbers $J=3/2$. To
this end, we consider the contribution of two baryons with masses $m_{1}$ and
$m_{2}$ into the polarization operator $\Pi^V(p,k)$
(\ref{pmn})
\bq
\frac{V_{\rho} < 0 \mid \eta_\mu \mid \Delta_{1} > <
\Delta_{1} \mid j_{\rho}^{em}
\mid \Delta_{2} > < \Delta_{2} \mid \bar{\eta}_\nu
\mid 0 >}{(p^{2} - m_{1}^{2})((p+k)^{2}
- m_{2}^{2})},
\label{15b}
\eq
where $\Delta_{1}$ and $\Delta_{2}$
are baryon states with masses $m_{1}$ and $m_{2}$
respectively.
Here we consider
 the case when  only spinor parts of the Rarita-Shwinger fields interact
with a photon.
In such case,
 the matrix element of the  electromagnetic  current  has
the following form:
\bq
\centerline{$< N_{1} \mid j_{\rho}^{em} \mid N_{2} > = $}
\nonumber
\\
\centerline{$\bar{v}^{(1)}_\mu (p)g_{\mu\nu} [ f_{12}(k^{2})
\gamma_{\rho} + \frac{\varphi_{12}(k^{2})}{m_{1}+m_{2}}\sigma_{\rho \lambda}
k_{\lambda}
+
\psi_{12}(k^{2})k_{\rho} ]v^{(2)}_\nu (p+k) =   $}
\nonumber
\\
\centerline{$\bar{v}^{(1)}_\mu (p) g_{\mu\nu}[ (f_{12}(k^{2}) +
\varphi_{12}(k^{2}))\gamma_{\rho} +
{\cal P}_{\rho}\frac{\varphi_{12}(k^{2})}{m_{1}+m_{2}} +
\psi_{12}(k^{2})k_{\rho} ] v^{(2)}_\nu (p+k)$}
\label{16b}
\\
\centerline{$< 0 \mid \eta_\mu \mid N, J^{P}=3/2^{+} > = \lambda v_\mu (p)$}
\nonumber
\\
\centerline{${\cal P}_{\mu} = p_{\mu} + (p+k)_{\mu}$,}
\nonumber
\eq
where $v_\mu (p)$ is a Rarita-Shwinger spin-vector
 satisfying the Dirac equation: $(\hat{p}-m)v_\mu (p)=0$,
$\gamma_\mu v_\mu =0$, $p_\mu v_\mu =0$.

Using (\ref{16b}) we can transform (\ref{15b}) to the following form:
\bq
\frac{\lambda_1\lambda_{2}}
{(p^{2} - m_{1}^{2})((p+k)^{2} - m_{2}^{2})}
V_{\mu} [g_{\mu\nu}\hat{p}_1\gamma_\rho\hat{p}_2 G_M(k^2)/3
 \label{17b}
\eq
$$
+(other\; structures\; with\;\gamma_\mu\; placed\; at\; the\; beginning\;
and\;\gamma_\nu\; at\; the \; end \; of\; them)]
$$
where $G_M$ is the magnetic form factor.

It is important to note that there
is no spin-$1/2$  baryon contribution in the structure $g_{\mu\nu}
\hat{p}_1\gamma_\rho\hat{p}_2$ which has the following amplitude:
$$
<0\mid\eta_\rho\mid J=1/2>=(Ap_\mu +B \gamma_\mu)u(p)
$$
where $(\hat{p}-m)u(p)=0$ and $Am+4B=0$.

{}From (\ref{17b}), it is obvious that the structure
$g_{\mu\nu}\hat{p}_{1}\gamma_{\rho}\hat{p}_{2}$ (where $p_{1}=p$ and
$p_{2}=p+k$)
contains magnetic transition only, since $G_{M}^{12}/3 = f_{12}(k^{2}) +
\varphi_{12}(k^{2})$, where $G_{M}^{12}$ is the magnetic form factor. So, we
shall further consider only the structure  $g_{\mu\nu}\hat{p}_{1}
\gamma_{\rho}
\hat{p}_{2}$.

Now let us discuss the factor $1/3$ which have appeared in eq.(\ref{17b}).
  Consider  interaction of spin-$3/2$
particle with the electromagnetic field:
\begin{equation}
a(\overline{\Psi} _{\mu}(P_{1})g_{\mu\nu}(P_{1}+P_{2})_{\rho}\Psi
_{\nu}(P_{2})+
ib\overline{\Psi} _{\mu}(\sigma_{\rho\lambda}g_{\mu\nu}/2+
2g_{\mu\rho}g_{\nu\lambda}
)\Psi _{\nu}F_{\rho\lambda}
                     \label{1a}
\end{equation}
where $\frac{1}{2}\sigma _{\rho\lambda}=[\gamma _{\rho}\gamma _{\lambda}]$,
$\Psi _{\mu}$ is Rarita-Shwinger spin-vector field (
$(\hat{P} -m)=0, \gamma _{\mu}\Psi _{\mu}=0$), $F_{\rho\lambda}=\partial
_{\rho}
A_{\lambda}-\partial _{\lambda}A_{\rho}$.
The first term of eq.(\ref{1a}) corresponds to the spin-independent part of
electromagnetic interaction of $\frac{3}{2}$-spin particle and the second one
describes the spin-dependent interaction. To express the value of the magnetic
moment (at $Q^2=0$)
let us consider the case when $A_{0}=0, P_{0}=m, F_{\rho\lambda}=
\delta _{\rho i}\delta_{\lambda j}\epsilon _{ijk}H_{k}$, where $H_{k}$ is
 magnetic field. Then we have
\begin{equation}
ib\overline{\Psi} _{m}(g_{mn}\sigma _{ij}\epsilon _{ijk}/2+2\epsilon
_{mnk})\Psi
 _{n}H_{k}
=b\overline{\Psi} _{m}(g_{mn}\Sigma _{k}+2i\epsilon _{mnk})\Psi _{n}H_{k}
                                                             \label{2a}
\end{equation}
where $\Sigma _{k}=diag(\sigma _{k},\sigma _{k})$
	Now we see that the operator $(\Sigma _{k}g_{mn}+2i\epsilon _{mnk})$ is
equal to $2S_{k}$ where $S_{k}$ is  spin operator for the Rarita-Shwinger
field.
So, the maximal energy of the particle in the magnetic field is equal to
\begin{equation}
E_{int.}=3bH=\mu H,
                                                            \label{3a}
\end{equation}
where $\mu$, by definition, is magnetic moment  or
magnetic form factor at $Q^2=0$.
Thus, we have
\begin{equation}
\mu =3b
\label{4a}
\end{equation}

Now let us consider the double dispersional relation for the function
at tensor structure $g_{\mu\nu}\hat{p}_1\gamma_\rho\hat{p}_2 V_\rho$:
\bq
f(P_{1}^{2}, P_{2}^{2}, Q^{2}) = \frac{1}{\pi^{2}} \int_{0}^{\infty}
\int_{0}^{\infty} \frac{\rho (s_{1}, s_{2}, Q^{2})}{(P_{1}^{2} +
s_{1})(P_{2}^{2} + s_{2})},
\label{18f}
\eq
where $P_{1}^{2}=-p_{1}^{2}$, $P_{2}^{2}=-p_{2}^{2}$, $Q^{2}=-k^{2}
\geq 0$ and $\rho(s_{1}, s_{2}, Q^{2})$ is the spectral density. Due to reasons
mentioned above, we cannot calculate $\rho(p_{1}, p_{2}, Q^{2})$ directly, but
we can consider the double Borel transformed structure function of correlator
(\ref{1}). Notice, that under replacement $p_{1} \rightarrow p_{1}+kz$, $p_{2}
\rightarrow p_{2}+kz$ the structure $\hat{p}_{1}\gamma_{\mu}\hat{p}_{2}
V_{\mu}$ transforms to
\bq
\hat{p}_{1}\gamma_{\mu}\hat{p}_{2}V_{\mu} \rightarrow (\hat{p}_{1}+
\hat{k}z)\hat{V}(\hat{p}_{2}+\hat{k}z) = \hat{p}_{1}\hat{V}\hat{p}_{2} -
2(Vp_{1})\hat{p}_{1}z -
\label{19b}
\\
2(Vp_{2})\hat{p}_{2}z + (p_{1}^{2} + p_{2}^{2})z\hat{V} + z^{2}(2(Vk)\hat{k}
- \hat{V}k^{2})
\nonumber
\eq
and all other structures in (\ref{17b}) with
 smaller number of $\gamma-$matrices near $g_{\mu\nu}$,
 could not  be transformed
 into $\hat{p}_{1}\hat{V}
\hat{p}_{2}$ . So we can extract
$g_{\mu\nu}\hat{p}_{1}\hat{V}\hat{p}_{2}$ in the
integral (\ref{1}). Then the integral
representation for the double Borel transformed structure function under
consideration is obtained from (\ref{18f}) by applying the operator $\hat{O}$:
\bq
\hat{O}f(P_{1}^{2}, P_{2}^{2}) = \int_{-\infty}^{+\infty} dz
\hat{\cal B}_{P_{1}^{2}}\hat{\cal B}_{P_{2}^{2}}f((P_{1}+kz)^{2},
(P_{2}+kz)^{2})
\label{20b}
\eq
where
$$
\hat{\cal B}=\lim_{n\rightarrow\infty}\frac{(P^2)^{n+1}}{n!}
\left( -\frac{\partial}{\partial P^2}\right)^n
$$
$$
P^2/n=M^2
$$
$M^2$ is the Borel parameter.

Applying the operator $\hat{O}$ to the left-hand and right-hand sides of
double dispersion relation (\ref{18f}) we get
\bq
f(M^{2}, Q^{2}) & = & \frac{1}{\pi^{2}}\int_{-1/2}^{+1/2}\exp
(\frac{Q^{2}z^{2}}{M^{2}}) \times
\nonumber
\\
\int_{0}^{\infty}\int_{0}^{\infty}ds_{1}ds_{2}\exp (-\frac{s_{1}+
s_{2}}{2M^{2}} & + & \frac{(s_{1}-s_{2})z}{M^{2}})\rho (s_{1},s_{2},Q^{2})
\label{21b}
\eq
where $M_{1}^{2}=M_{2}^{2}=2M^{2}$.

We shall use the sum rules (\ref{21b}) to calculate the nucleon form factor.
The function $f(M^{2},Q^{2})$ is calculated using operator expansion in the
external field, and the spectral density $\rho (s_{1},s_{2},Q^{2})$
is saturated by intermediate states with quantum numbers of the current
(\ref{eta}).

There are two different types of intermediate state contributions into
(\ref{21b}). The first is responsible for the diagonal transitions between the
states with equal masses. The second is responsible for nondiagonal transitions
between the states with different masses. In the first case the right-hand
side of (\ref{21b}) obviously has the form
\be
\lambda^{2}\frac{G_{M}}{3}(Q^{2})e^{-m^{2}/M^{2}}\int_{-1/2}^{+1/2}
e^{\frac{Q^{2}z^{2}}{M^{2}}}dz
\label{22b}
\ee
where $\lambda^{2}$ is the square of the residue of the state with mass $m$
into
the current $\eta_\mu$ defined by formula (\ref{16b}), $G_{M}(Q^{2})$ is the
corresponding magnetic form factor.

In the second case for the transition between states with masses $m_{1}$ and
 $m_{2}$ we get:
\bq
\lambda_{1}\lambda_{2}\frac{G_{M}^{(12)}}{3}
(Q^{2})e^{-\frac{m^{2}_{1}+m^{2}_{2}}{2M^{2}}}
\int_{-1/2}^{+1/2}e^{\frac{Q^{2}z^{2}+(m^{2}_{1}-m^{2}_{2})z}{M^{2}}}dz
\label{23b}
\eq
Now, to investigate the properties of (\ref{22b}) and (\ref{23b}) let us
expand them in the series on $Q^{2}/M^{2}$
\bq
\lambda^{2}\frac{G_{M}(Q^{2})}{3}
e^{-m^{2}/M^{2}}(1 + \frac{1}{12}\frac{Q^{2}}{M^{2}} +
\frac{1}{5\cdot 2^{5}}(\frac{q^{2}}{M^{2}})^{2} + ...\;)
\label{24b}
\eq
\bq
\beta_{1}\beta_{2}G_{M}^{12}(Q^{2})e^{-\frac{m^{2}_{1}  +  m^{2}_{2}}{2M^{2}}}
\frac{2M^{2}}{m_{1}^{2}-m_{2}^{2}}
 \sinh (\frac{m_{1}^{2}-m_{2}^{2}}{2M^{2}})\times
\label{25b}
\\
\{1 + \frac{Q^{2}}{M^{2}}[\frac{1}{4}-\frac{M^{2}}{m_{1}^{2}-m_{2}^{2}}
\coth(\frac{m_{1}^{2}-m_{2}^{2}}{2M^{2}})  +
 \frac{2M^{4}}{(m_{1}^{2}-m_{2}^{2})^{2}}] + ...\}
\nonumber
\eq
{}From (\ref{24b}) and (\ref{25b}) we see that  diagonal
 transitions  of the excited states
 are exponentially suppressed   compared to the $\Delta$-isobar
contribution in
(\ref{24b}).
 Let us write the non-suppressed part of the contribution from the
nondiagonal transition between the nucleon with mass $m_{N}$
  and an excited state
with mass $m_{\Delta^{*}}$
\bq
\lambda\lambda_{\Delta^{*}}\frac{G_{M}^{\Delta\Delta^{*}}(Q^{2})}{3}
\frac{e^{-\frac{m^{2}_{\Delta}}{M^{2}}}
M^{2}}{m_{\Delta^{*}}^{2} - m_{\Delta}^{2}}\{1 & + &
\nonumber
\\
\frac{Q^{2}}{M^{2}}[\frac{1}{4} - \frac{M^{2}}{m_{\Delta^{*}}^{2}-
m_{\Delta}^{2}}\coth
(\frac{m_{\Delta^{*}}^{2}-m_{\Delta}^{2}}{2M^{2}}) & + & \frac{2M^{4}}
{(m_{\Delta^{*}}^{2}-m_{\Delta}^{2})^{2}}] + ...\}
\label{26b}
\eq
where $m_{\Delta}$, $m_{\Delta^{*}}$, $\lambda$ and
$\lambda_{\Delta^{*}}$ are masses
 and residues of  $\Delta$-isobar and its resonance
$\Delta^{*}$ respectively.

Expression (\ref{26b}) is analogous to the contribution from the
 single-pole term
    appearing in QCD sum rules for correlators in  constant external
 field (see
   \cite{ext}).

It is easy to see that the function multiplied by $Q^{2}/M^{2}$ in (\ref{26b})
 changes from $1/12$ at $m_{2}^{2} - m_{\Delta}^{2} \rightarrow 0$ to
 $1/4$ at $m_{2}^{2} - m_{\Delta}^{2} \rightarrow \infty$.
However, taking into account the continuum, only contributions from the
 states with $m_{2}^{2} - m_{\Delta}^{2} \sim s_{0} - m_{\Delta}^{2}$
($s_{0}$ is
 the continuum threshold) are to be considered.
 In the region $s \gg s_{0}$, our model of
    continuum is quite correct, but when $s \sim s_{0}$ it is not so.
 Then we see, that nonexponentially suppressed terms will be
determined by the states with $m_{2}^{2} \sim s_{0}$.
Taking $m_{2}^{2}-m_{\Delta}^{2} \simeq s_{0}-m_{\Delta}^{2}
 \simeq 1.5 GeV^{2}$, $M^{2} \sim 1 GeV^{2}$, expression (\ref{26b}) can be
 written in the form
\bq
\lambda_\Delta\lambda_{\Delta *}
\frac{G_{M}^{\Delta\Delta^*}(Q^{2})}{3}
\frac{e^{-\frac{m^{2}_{\Delta}}{M^{2}}}
M^{2}}{m_{\Delta^{*}}^{2} - m_{\Delta}^{2}}(1 +
\frac{Q^{2}}{12M^{2}}(1+\epsilon)+...\;)
\label{27b}
\eq
where $\epsilon < 0.1$.

Thus, from (\ref{24b}) and (\ref{27b}) it is seen that when $Q^{2}/M^{2} \leq
1$,
   and $M^{2} \sim 1 GeV^{2}$, the right-hand side of (\ref{21b}) is
\bq
\frac{1}{3}
\lambda_\Delta^{2}e^{-m_{\Delta}^{2}/M^{2}}(G_{M}(Q^{2}) + C(Q^{2})M^{2})
\int_{-1/2}^{+1/2}e^{\frac{Q^{2}z^{2}}{M^{2}}}dz
\label{28b}
\eq
The accuracy of (\ref{28b}) is of an order of a few percents (it
depends on the numerical value of
$\epsilon$ from (\ref{27b})). It can be shown, that the next terms in
the expansion
(\ref{25b}) in powers of $Q^{2}/M^{2}$ do not change the situation.
 So, in the region $Q^{2}/M^{2} \leq 1$, $M^{2} \sim 1 GeV^{2}$, the
 right side of the sum rule
(\ref{21b}) is indeed represented by expression (\ref{28b}).

Let us note that when $Q^{2}/M^{2} > 1$, the expression (\ref{28b}) is
 invalid and we cannot separate the contribution of the
 single-pole terms from the contribution of the us
 double-pole term $\sim G_{M}(Q^{2})$ which we are interested in.
Thus, our sum rule is expected to be
 valid in the region $0 \leq Q^{2} < 1 GeV^{2}$.

Here we have constructed the right-hand "phenomenological" side
 of the sum rule
(\ref{21b}) and discussed the region of its applicability. Now, let us pass
to the calculation of the left, "theoretical" part of our sum rule.

Using eq.(\ref{6}) and dispersion integral (\ref{21b}) we have
obtained the following sum rule for $G_M$:
$$
\int_{-1/2}^{+1/2}e^{Q^2z^2/M^2}dz\int_0^{s_0}ds_1\int_0^{s_0}ds_2
e^{-\frac{s_1+s_2}{2M^2}+z\frac{s_1-s_2}{M^2}}\frac{\rho_{\Delta}
(s_1,s_2,Q^2)}{\pi^2}
$$
\be
+\frac{2}{3}a^2\int_{-1/2}^{+1/2}dz e^{Q^2z^2/M^2}
\label{sumrule}
\ee
$$
+\frac{2}{3}(2\pi)^2a(i\Pi_2(Q^2))M^2e^{\frac{Q^2}{4M^2}}
\nonumber
$$
$$
-\frac{2}{9}a^2(1+Q^2(i\Pi_4(Q^2)))e^{\frac{Q^2}{4M^2}}
$$
$$
-(2\pi)^2\frac{Q^2}{3}\Pi_1(Q^2)e^{\frac{Q^2}{4M^2}}
=I(M^2,s_0,Q^2)\int_{-1/2}^{+1/2}e^{\frac{Q^2z^2}{M^2}}dz
$$
$$
=\tilde{\lambda}_{\Delta}^2e^{\frac{-m_\Delta^2}{M^2}}e_u^{-1}(
\frac{1}{3}G_M(Q^2)+C(Q^2)M^2
\int_{-1/2}^{+1/2}e^{\frac{Q^2z^2}{M^2}}dz
$$
where $e_u=+\frac{2}{3}$ and
$\rho_\Delta (s_1,s_2,Q^2)$ is a spectral
density:
$$
\frac{1}{\pi^2}\rho_\Delta (s_1,s_2,Q^2) =
Q^2\frac{(\kappa-U)^2}{2^6\kappa^3}
\left(\frac{8}{3}(2\kappa +U)+Q^2(1+
\frac{U(2\kappa +U)}{\kappa^2})\right)
$$
where
$
\kappa = \sqrt{(s_1+s_2+Q^2)-4s_1s_2}
$,
$
U=s_1+s_2+Q^2
$,
$
a=(2\pi )^2\mid <\bar{\psi}\psi >_0\mid =0.55GeV^3
$,
$\tilde{\lambda}_\Delta =(2\pi )^2\lambda_\Delta = 2.3GeV^6
$,
$
s_0=4GeV^2
$,
$
m_\Delta =1.232GeV
$,
$$
i\Pi_2(Q^2)=-<\bar{\psi}\psi>_0\left(
\frac{4}{Q^2+0.6}-\frac{2}{Q^2+2.6}\right)
(GeV)
$$
Here $Q^2$ is in units of $GeV^2$.
$$
\Pi_1(Q^2)=\frac{1}{(2\pi)^2}
\left( -\frac{2m_\rho^2(2\pi)^2}{g_\rho^2(Q^2+m_\rho^2)}+\ln
(1+\frac{s_0^*}{Q^2}
\right),
$$
where $\frac{g_\rho^2}{4\pi}=2.3$ is  residue of the $\rho-$meson,
$m_\rho^2=0.6GeV^2$ is the $\rho-$meson mass squared, and
$s_0^*=1.5GeV^2$. Numerically in GeV units we have
$$
\Pi_1(Q^2)=\frac{1}{(2\pi)^2}
\left(-\frac{1.5}{Q^2+0.6}+\ln (1+\frac{1.5}{Q^2})\right)
$$
$$
\Pi_4(Q^2)=\frac{1}{2}\Pi_2(Q^2)
$$
All these expressions for operators $\Pi_i(Q^2)$ were obtained in
\cite{fn}.
In (\ref{sumrule}
$s_{0} = 4 \,\,GeV^{2}$ is continuum threshold, which is numerically equal
to
the threshold of continuum in the sum rules for $\Delta$-isobar mass.
Such a choice for
the value of the threshold follows from the assumption that
there is one and the same value
of threshold for different structures in the sum rules. Then,  considering
the sum rules for the electric charge (i.e. for the electric form factor
$G_{E}(Q^{2})=1$,  $Q^{2}=0$)  it  can  be  shown  that  they
coincide with the mass
sum rules except for the possibly different values of thresholds. But from
Ward identity, we know that these sum rules should coincide exactly. So the
value of the threshold in the sum rules is $s_{0} \simeq 4 GeV^{2}$
\cite{baryon}. This assumption is based on the physical meaning of the
quantity
$s_{0}$, which  as we know from the experience is determined mainly by the
position of the next resonance in corresponding channel. And it would be
very surprising if the resonance contribution to the electric form factor's
channel strongly differs from the contribution to the magnetic form factor's
channel.

In the limit $s_0\rightarrow\infty$
\be
\frac{1}{\pi^2}\int_0^\infty ds_1\int_0^\infty ds2
\rho_\Delta (s_1,s_2,Q^2) e^{-\frac{s_1+s_2}{2M^2}+z\frac{s_1-s_2}{M^2}}
\label{ad}
\ee
$$
=e_uM^6\left( 2E_2(\frac{Q^2}{4M^2}(1-4z^2))-3E_3(frac{Q^2}{4M^2}(1-4z^2))
+E_5(\frac{Q^2}{4M^2}(1-4z^2)) \right)
$$
Here $E_n(z)=\int_1^\infty x^{-n}e^{-zx}dx$.

Now, let us discuss the  continuum  contribution into the second term of the
left side of (\ref{sumrule}). The double discontinuity of the
corresponding diagram could not
be calculated due to the reasons discussed in the Section 2
of this paper. We do not know the exact formula, analogous to the first
term in (\ref{sumrule}), but can write an approximate expression, which becomes
exact in the limit when $M^{2} \rightarrow \infty$. Moreover, if one
substitutes the first term in the left side of (\ref{sumrule}) by the analogous
term

\bq
 e_{u} M^{6} (1-e^{-s_{0}/M^{2}} (1 +
\frac{s_{0}}{M^{2}} +
\frac{s_{0}^{2}}{2M^{4}})) e^{Q^{2}/4M^{2}} \times
\label{39}
\\
\int _{-1/2}^{+1/2}
dz [ 2E_{2}(\frac{Q^{2}}{4M^{2}}(1-4z^{2})) -
3 E_{3}(\frac{Q^{2}}{4M^{2}}(1-4z^{2})) +
E_{5}(\frac{Q^{2}}{4M^{2}}(1-4z^{2}))]
\nonumber
\eq
then one sees that their differences is less then $10\%$ in the region
 $Q^{2} < 1 GeV^{2}$. So, since we shall work in such a region of
$M^{2}$, where
 the continuum contribution is less then $20-30\%$, then the uncertainties
due to the second term in (\ref{sumrule}) will be less then a few percents.

To find $G_M(Q^2)$
we should study the following formula to kill nonsuppressed contribution
of the nondiagonal transitions
$$
G_M(Q^2)=3e_u(1-M^2\frac{\partial}{\partial M^2})\tilde{\lambda}^{-2}
e^{m_\Delta^2/M^2}
I(M^2,S_0,Q^2)
$$
at any fixed $Q^2$.

The results obtained for the form factor are depicted in Fig.1.
Notice that contribution of nondiagonal transitions is very small
(about few percents) in the sum rule. In the region $0\leq Q^2\leq 0.8
GeV^2$ the result obtained may be fitted by the following relation
\be
G_M(Q^2)=\frac{6.16}{1+Q^2/\mu^2}\; \left(\frac{eh}{2m_pc}\right)
\label{fit}
\ee
where $\mu^2=0.7\;GeV^2$.

The additive quark model prediction is $G_M(0)=7.4\frac{eh}{2m_pc}$.

The analogous sum rule could be also written in the case of
$\Omega^-$-hyperon in the limit of $SU(3)$ symmetry after
evident interchange $e_{u} \leftrightarrow e_{s}$.
An accuracy of the results obtained is $10-20\%$.

\section{Conclusions}

In the present paper the QCD sum rules for the polarization operator in
a variable external field are used to calculate
the $\Delta$-isobar magnetic form factor in the infra-red $0 \leq Q^{2} <
1 GeV^{2}$ region. Analogous sum rules can be used for calculation of other
diagonal hadronic form factors in the infra-red region.
 Notice that this method does not work in the case of nondiagonal
form factors because it is not possible  to separate interesting contribution
 into correlator. There is  only one way to
do it: to sum all $(kx)^n$-terms in the expression for
nonperturbative propagator (\ref{5}). It was done only
in the case $Q^2=0$ \cite{braun} using operator product expansion
on a light cone and models for the photon wave functions.

{\bf Acknowledgments}

I am grateful to the Theoretical Division of Max-Plank-Institute
in Munich, where this paper was started, and CEBAF
Theory
Group, where it was completed, for warm
hospitality and interesting discussions.

\end{document}